\documentclass[aps,prl,twocolumn,nofootinbib]{revtex4}

 \usepackage{amsmath}
 \usepackage{graphicx}
\usepackage{relsize}
\begin{document}

\title {A method to obtain bounds on the equation of state of cold nuclear matter from imaginary chemical potentials}

\author{Thomas D. Cohen}
\email{cohen@umd.edu}

\affiliation{Department of Physics, University of Maryland,
College Park, MD 20742-4111}

\begin{abstract}
The sign problem in numerical calculations of the QCD Euclidean space path integral of QCD with a chemical potential vanishes if the chemical potential is imaginary.  Moreover, calculations of the partition function with imaginary chemical potentials are equivalent to calculations with  Lagrange multipliers enforcing the current density.  At zero temperature, Lorentz boosts allow one to deduce properties of systems with both number density and current density from properties of systems with a current density alone; this allows both upper and lower bounds to be determined for the equation of state (EOS) in the form of energy density as a function of number density.
\end{abstract}
\maketitle

Ever since the semi-empirical mass formula of von Weizs\"acker\cite{vW} was developed nearly a century ago, the properties of cold infinite nuclear matter have played a central role in nuclear physics.  Ideally, these properties could be understood directly from QCD.   In the case of cold neutron matter, the need for {\it ab initio} QCD calculations  for the equation of state (EOS) is particularly important; it is a key input to calculations of neutron star properties  but is not well constrained from terrestrial experiments (for recent reviews see refs.~\cite{NS1,NS2}).

Unfortunately, calculations of the EOS are not possible using conventional lattice methods  since the inclusion of a chemical potential introduces an intractable sign problem \cite{SP0,SP1,SP2}.  While various methods have been proposed to evade or mitigate sign problems (for recent reviews see refs.~\cite{SPsol,SPsol0,Spsol1,SpSOl2}), at present none are robust enough to make QCD at nonzero chemical potential viable.  In the future, it is conceivable that quantum computers might provide a way forward but the treatment of quantum field theories on quantum computers is in its infancy \cite{QC1,QC2,QC3,QC4,QC5,QC6,QC7,QC8,QC9} and  sufficiently large and quantum computers with  sufficiently low noise to attack the problem in a meaningful way is likely to be far in the future.

This paper focuses on a much less ambitious goal: rather than attempting to evade the sign problem to directly calculate the EOS, the goal is to evade the sign problem and provide meaningful constraints on the EOS of cold matter---both upper and lower bounds.  The possibility of constraining the equation of state without incurring a sign problem is not new.  More than two decades ago\cite{Cohen:2003ut}, it was noted that the basic method of QCD inequalities (for a review see ref.~\cite{Nussinov:1999sx}) can do this: an isospin chemical potential in QCD with two  degenerate light flavors has a functional determinant which is manifestly positive \cite{Kogut:2002zg,Son:2000xc} and lacks a sign problem while the determinant for a baryon chemical potential has the same magnitude and a fluctuating phase which implies that the free energy of isospin matter provides an upper bound for the free energy of isospin symmetric nuclear matter, and can do so at any temperature.  More recently, this approach was extended more generally to phase-quenched evaluation of the QCD functional integral \cite{Moore:2023glb} and numerical simulations \cite{Abbott:2024vhj}, and constraints have been obtained at nonzero temperature \cite{Fujimoto:2023unl}.  Such methods are interesting but have important limitations: they provide upper bounds only and not lower bounds; even as upper, bounds they are likely to be very weak at low and modest densities and low temperatures.  More generally, it is useful to have multiple sources of constraints for the EOS; taken together, they may well do a better job in constraining the EOS.

This paper introduces a novel way to constrain the EOS for cold matter while evading the sign problem. The approach provides distinct upper and lower bounds for the energy density as a function of the baryon number density, $\epsilon (n)$. These constraints can be deduced entirely from calculations of QCD at imaginary chemical potential, provided that such calculations are viable at sufficiently low temperatures to extrapolate to  $T=0$. 

Of course, analytic continuation from an imaginary chemical potential has long been a method to obtain information about the EOS at high temperatures \cite{DEFORCRAND2002290} and has been refined substantially over the years (for a review see ref.~\cite{Guenther:2020jwe}); however, that technique breaks down at low temperatures.  Here, a distinct method is proposed that is valid for $T=0$ (and only $T=0$) but only gives bounds on the EOS rather than the EOS itself.   

The analysis underlying these bounds holds for theories well beyond QCD; it applies to a wide class of quantum field theories. The discussion here focuses on QCD as a paradigmatic example.   For simplicity of notation, the analysis is done for a single chemical potential coupled to baryon number, but the generalization to separate chemical potentials for distinct flavors is straightforward.  

The bounds for $\epsilon(n)$ at $T=0$ are
\begin{subequations}
\begin{align}
\epsilon(n_\beta^u ) & \le \frac{ T_{zz}^E +  \beta^2 \, T_{tt}^E}{1-\beta^2} \; \; {\rm with}  \; \; n_\beta^u  \equiv \frac{  \beta \,  n^E}{\sqrt{1-\beta^2}} \label{Eq:ub}
\\
\epsilon(n_\beta^l) & \ge \frac{1- \beta^2}{1+ \beta^2 \frac{P(n_\beta^l)}{\epsilon(n_\beta^l)}} T_{zz}^E \ge \frac{1- \beta^2}{1+ \beta^2 } T_{zz}^E \label{Eq:lb} \\
 {\rm with} & \;\;  n_\beta^l  \equiv \frac{ \sqrt{1-\beta^2} \,  n^E}{\beta} \nonumber
\end{align}
\end{subequations}
where the superscript $E$  denotes values obtained from the Euclidean-space functional integral;  $T_{tt}^E$, $T_{zz}^E$ and $n^E$, are the energy density, pressure, and baryon number density, calculated with a Lagrangian multiplier $\mu^E$ that fixes $n^E$.  Given the conversion from Minkowski to Euclidean space, a Euclidean chemical potential $\mu^E$ corresponds to an imaginary chemical potential of the same magnitude in Minkowsky space.  These inequalities hold
for all values of a real parameter $\beta$ with $-1 < \beta < 1$.
Inequality~(\ref{Eq:lb}) has two forms, a tighter bound that depends on $ \frac{P(n_\beta)}{\epsilon(n_\beta)}$ and looser one that takes $ \frac{P(n_\beta)}{\epsilon(n_\beta)}$ to be unity, its maximum possible value consistent with causality.   The tighter form depends on $ \frac{P(n_\beta)}{\epsilon(n_\beta)}$, which is typically unknown.  However, the form remains valuable since any proposed model for the EOS (which automatically fixes $\frac{P(n_\beta)}{\epsilon(n_\beta)}$) must be consistent with the tighter bound.

Inequalities (\ref{Eq:ub}) and (\ref{Eq:lb}) depend on an arbitrary parameter $\beta$.  Starting from calculations with a single fixed $\mu^E$ (and hence fixed $n^E$), sweeping through $\beta$ sweeps through all $n$, providing upper and lower bounds.   However, more stringent bounds for any density can be found by calculating at multiple values of $\mu^E$; for each fixed value of $\beta$ corresponding to the upper (lower) bound for the density of interest, one chooses the smallest (largest) upper (lower) bound among these.  In general, as the limits $\beta \rightarrow 0$ or $\beta \rightarrow \pm 1$ are approached, the bounds become increasingly weak and, in the limit, useless.  $\beta^2 = \frac12$, is special; starting from the same value of $n^E$, it gives $n_\beta^l=n_\beta^u$ yielding upper and lower bounds for the EOS evaluated at the same density.

The derivation is straightforward: the analysis depends on $Z^{E}(\mu^E)$, the partition function  being  extractable from the Euclidean space QCD functional integral  with a Lagrange multiplier $\mu^E$ for the baryon density (corresponding to an imaginary chemical potential of $-i \mu^E$ in Minkowsky space) with sufficient accuracy to extract
\begin{equation}
    \Omega(\mu^E ) \equiv \lim_{\mbox{\tiny $\begin{matrix}
   \beta \rightarrow \infty\\L_x\rightarrow \infty \\L_y\rightarrow \infty\\L_z\rightarrow \infty
   \end{matrix}$}}
    \log \left(\frac{ Z^{E}(\mu^E)}{Z^{E}(0)}\right ) \label{Eq:Omega}
\end{equation}
for a rectangular Euclidean box of temporal size $\beta$ and spatial sizes of $L_x,L_y,L_z$.
In practice, this would need to be done via lattice Monte Carlo methods and would, at best, only be approximately true, given statistical limitations and extrapolations from finite volumes and lattice spacing.  The key assumption is that there are no technical impediments to such an extraction at sufficient accuracy to yield meaningful results; a central point is that there is no sign problem for such calculations.  

While imaginary chemical potentials are not physical, they directly yield physical observables---albeit ones that are not typically studied.  Consider a Euclidean-space QCD functional integral with a Lagrange multiplier, $\lambda_z$, that enforces a fixed average value for the baryon current density (which for concreteness will be taken to be in the $\hat{z}$ direction), rather than the usual chemical potential: $
    {\cal L} =     {\cal L}_{\rm QCD} - \lambda_z j_z $ with $ j_z  \equiv \overline q \gamma_z q$.
Given the connection between Euclidean space and Minkowski space, the generating functions  for QCD with a current density fixed by $\lambda_z$ and for QCD  with a number density fixed by $\mu$ analytically continued to  imaginary $\mu$, are related by
$Z_{j}^{E}(\lambda_z) =  Z^{E}(\mu^E)$ where $Z_{j}^{E}(\lambda_z)$ is the partition function including a Lagrange multiplier for baryon current density. Combining this with standard thermodynamic arguments implies that at $T=0$
\begin{equation}
\begin{split}
\Omega(\lambda_z) & \equiv T_{tt}^E(\lambda_z) - (\lambda_z) j_z =  - T_{zz}(\lambda_z)  \; \; {\rm with} \\ j_z(\lambda_z)  &= \frac{\partial \, \Omega(\lambda_z) }{\partial \, \lambda_z}
\label{Eq:connection}
\end{split}\end{equation}
 where $\Omega(\lambda_z)$ is given in Eq.~(\ref{Eq:Omega}) evaluated at $\mu^E= \lambda_z$.  {As an aside, the system is not isotropic, resulting in unusual thermodynamic properties.}  The key point is that $T_{tt}^E(\lambda_z)$, $T_{zz}^E(\lambda_z)$, and $j_z(\lambda_z)$ are all extractable from calculations with an imaginary chemical potential.

Access to $T_{tt}(\lambda_z)$,  $T_{zz}(\lambda_z)$ and $j_z(\lambda_z)$ allows one to deduce their values in a frame boosted in the $\hat z$ direction by $\beta$ since  $T_{\mu \nu}$ is a 4-tensor and $n_\mu$ is a 4-vector:
\begin{equation}
\begin {split}
    T'_{tt} &= \frac{T_{tt} + \beta^2 T_{zz}}{1-\beta^2}  \;, \; 
    n' = \frac{\beta \,  j_z}{\sqrt{1-\beta^2}}  \; , \;  j_z'= \frac{ j_z}{\sqrt{1-\beta^2}}  
    \end{split}
\end{equation}
where the prime indicates the boosted frame.  Since the system with $\lambda_z$ is physical, the boosted system is as well. Moreover, boosting a zero temperature system other than the vacuum, yields a distinct  $T=0$ system.  This system has an energy density $T'_{tt}$ and baryon number density $n'$.  On the other hand, $\epsilon(n)$ is the lowest energy density for a given value of $n$, so  $\epsilon(n') \le T'_{tt}$.  Combined with Eq.~(\ref{Eq:connection}), this immediately yields the upper bound given in inequality (\ref{Eq:ub}).

Inequality (\ref{Eq:lb}) can be derived similarly; consider a system with fixed baryon density and boost it to obtain a system with nonzero current density; this yields an upper bound for $T_{zz}^E$ which implies a lower bound for $\epsilon$.  The boosted value of   $T_{zz}^E$ depends on both the energy density and pressure of the system with fixed baryon density, so the lower bound depends on the ratio $\frac{P}{\epsilon}$ at the baryon density of interest.

\begin{figure*}[t]
    \includegraphics[width=.8
\textwidth]{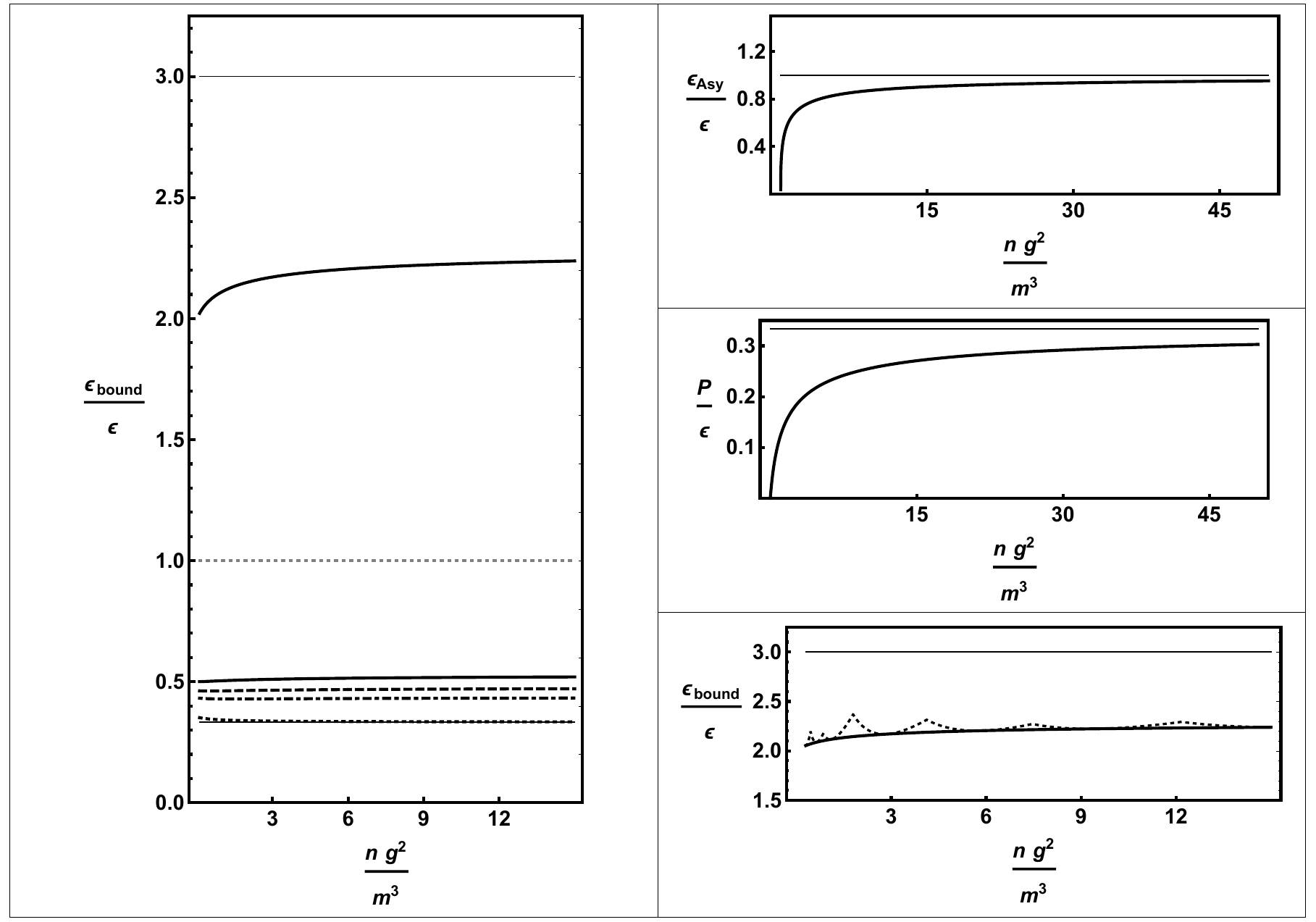}
    \caption{Bounds from the classical $\phi^4$  theory discussed in the text. The left hand panel shows the ratio of the worst case upper bound (thin line) and optimal upper bound obtainable from Eq.~ (\ref{Eq:ub}) (thick line) energy density to the actual energy density as a function of density above unity and below unity shows the analogous results for the lower bound;  the lower bound depends on $P/\epsilon$, so several values of  $P/\epsilon$ are shown: zero (thick line), 1/6 (dashed line), 1/3 (dot-dashed line) and 1 (dotted line); the worst case is shown by a thin line which is almost overlapped by the dotted line of $P/\epsilon$=1, except at very low $n$.  The left panel covers $n$ up to 15 since the energy is approaching its asymptotic value by then (as seen in the upper right-hand panel).  The thick line middle right-hand panel shows the actual values of $P/\epsilon$, which may help interpret the left-hand panel (with the thin line indicating $1/3$), while the bottom right panel indicates the extent to which a handful of measurements allows one to approximate the optimal bounds--focusing on the upper bound.  In that figure, effects of measurements at $\frac{j_z g^2}{m^3} =.3, .6, 1.0 , 2.5, 5.0, 8.0$ and $13.0$ were simulated. }
    \label{fig:ClassFig}
\centering 
\end{figure*}

Lattice calculations of these bounds could use substantial computational resources; whether they are worth the cost presumably depends on the extent to which these bounds provide useful constraints. A worst-case analysis could provide some guidance
\begin{equation}
\begin{split}
   &\frac{ T_{zz}^E (\mu^E)}{3} \le \epsilon_{lb}(n)  \le \epsilon(n) \le   \epsilon_{ub}(n) \le 3 T_{zz}^E (\mu^E) \\
   &{\rm where} \; \; n = n^E (\mu^E) \; ;
   \end{split} 
\end{equation}
 $\epsilon_{ub}(n) $ and  $\epsilon_{lb}(n) $ are the strongest upper and lower bounds obtainable from Eqs.~(\ref{Eq:ub}) and (\ref{Eq:lb}) using multiple values of $\beta$ and $\mu^E$.  The worst case values of $\frac{ T_{zz} (\lambda_z )}{3}$ and $3 T_{zz} (\lambda_z)$ follow by taking $\beta=\frac12$ for both the upper and lower bounds and exploiting a causality constraint on the speed of sound that ensures that the maximum value that $P$ can attain when a chemical potential is present is $\epsilon$ and an analogous constraint for QCD with a Langrange multiplier for current density. At worst, there is less than an order of magnitude between the lower and upper bounds; the best upper and lower bounds obtainable from calculations with imaginary chemical potentials are likely to be substantially tighter than this.

One way to assess the extent to which the constraints are useful in constraining the EOS (beyond the information in the generic worst case analysis) is by studying tractable systems where the EOS and the bounds are separately calculable. There is no guarantee that QCD will track the behavior of any given tractable model, so it is useful to study a wide range of models to see if there is some generic behavior shared by multiple models.

As a first example consider  a classical $\phi^4$ bosonic field theory in 3+1 dimensions  with a single complex  $\phi$ field; the Lagrangian is ${\cal L} = \partial^\mu \phi^* \partial_\mu \phi  - m^2 \phi^* \phi  - g^2 \left (\phi^* \phi\right )^2$  and the $U(1)$ current density is $J_\mu =i\left ( \phi^* \partial_\mu \phi -(\partial_\mu \phi^* ) \phi \right )$.  Note that the derivation of the bounds of Eqs.~ (\ref{Eq:ub}) and (\ref{Eq:lb}) depends on relativistic transformations, which apply equally to classical and quantum systems.  This simple model has only one dimensional parameter, $m$, and one dimensionless parameter, $g$, allowing a scaling so that $\epsilon/(m^4 g^2)$ as a function of $n/(m^3 g^2)$ is independent of both $m$ and $g$.  At asymptotically large densities, the energy density scales with number density as  $\frac{\epsilon_{asy}}{m^4 g^2} = \frac34 \left ( \frac{n}{m^3 g^2}\right)^{\frac43}$. Thus, this system shares at least one important property with QCD, asymptotic scaling of $\epsilon$ as  $ n^{\frac 43}$.  The bounds for this simple model are shown in Fig.~\ref{fig:ClassFig}.  Minimally, Fig.~\ref{fig:ClassFig} shows that the calculable bounds are substantial improvements on the worst-case bounds.   

Future studies could be made for a wide variety of tractable relativistic field theory models in order to gain insights into the utility of these bounds.   Classical field theories that emulate various known features of QCD are likely to be instructive.  For example, classical bosonic field theories can include multiple global $U(1)$  charges (one for each dynamically relevant quark flavor) or they can build in the analog of nuclear matter saturation--{\it i.e.} zero pressure with nonzero baryon density.   More generally, it is likely to be instructive to study classical models that reproduce what is known phenomenologically about both the low density EOS (saturation density and binding energy and the incompressibility) as well as the high density $n^{\frac 43}$ scaling at high density.  Another class of tractable model worth studying are mean-field models of fermions in 3+1 dimensions.  

It is also sensible to study tractable quantum models with conserved $U(1)$ charges.  There are examples of such models in 1+1 dimension such as the massive Thiring (or equivalently the sine-Gordon \cite{Coleman:1974bu,Mandelstam:1975hb}) model that can be studied via analytic methods in both the weak and strong coupling limits; in intermediate coupling the finite density system can be solved and numerically via the Bethe ansatz \cite{Haldane1982,Bergknoff:1978wr,Bergknoff:1978bm}.  Other 1+1-dimensional models are tractable at finite density \cite{PhysRevD.62.096002} when studied in a large $N_c$ limit, including the 't Hooft model \cite{tHooft:1974pnl} and the Gross-Neveu model \cite{Gross:1974jv}.  
Finally, it would be helpful to study a real quantum field theory in  3+1 dimensions.  It turns out QCD itself can provide a test case.  Of course,   there remains no practical way to calculate the $T=0$ EOS as a function of baryon density (and if there were, the bounds would be irrelevant).  However, the bounds exist for any conserved $U(1)$ charge,  including the third component of isospin.  Moreover, in the regime where the pion mass, along with $\mu_I$ and $\lambda_I$ (the Lagrange multipliers for the isospin density and its current) are well below the characteristic hadronic scale, the energy density of QCD is well approximated by cacluations using the chiral Lagrangian at low order using the approach of ref.~\cite{Son:2000xc}; this allows for a meaningful test of the utility of the bounds for a 3+1 dimensional quantum field theory, albeit in a limited regime.  

Testing these bounds for a wide variety of models has a potential benefit beyond merely determining how constraining they turn out to be.  Such tests may also give at least a sense of whether a qualitative understanding of the actual EOS (as opposed to mere bounds) might be extracted.  One expects the true EOS to be well above the worst-case lower bound and well below the worst-case upper bound.  The simple classical $\phi^4$ model shown in Fig.~\ref{fig:ClassFig} has a remarkable property: the geometric mean of the worst-case upper and lower bounds for the energy precisely yields the true $T=0$ EOS for all $n$.  While that behavior is presumably an artifact of the extreme simplicity of the model, it is not totally implausible that the geometric mean of the worst-case upper and lower bounds generically tracks the actual EOS reasonably well.  It is worth testing the extent to which this is true by testing this for various models. If this appears to be generically true, it would greatly increase the value of studies of QCD with imaginary chemical potentials at low $T$.

Finally, it is worth noting that while there are no sign problems for imaginary chemical potentials, it is conceivable that unanticipated numerical difficulties could arise, given the lack of experience with numerical lattice calculations at very low temperatures with imaginary chemical potentials.  Thus, it would be helpful to do numerical studies of this sort in relatively simple situations (in lower dimensions and with a single quark flavor and perhaps with relatively heavy quark masses) before a serious attempt is made on QCD in its full glory.

\begin{acknowledgments}

This work was supported in part by the U.S. Department of Energy, Office of Nuclear Physics under Award Number(s) DE-SC0021143, and DE-FG02-93ER40762.

\end{acknowledgments}

 \bibliography{main.bib}
\end{document}